\documentclass{article}[11pt]

\usepackage[english]{babel}

\usepackage[a4paper,top=3.5cm,bottom=3.5cm,left=3cm,right=3cm,marginparwidth=2cm]{geometry}

\usepackage{amssymb} \usepackage{color,graphicx} \usepackage{amsmath}
\usepackage{amsbsy} \usepackage{amsthm} \usepackage{wrapfig}
 \usepackage{float} \usepackage{braket}
\usepackage{placeins}
\usepackage{comment,subfigure}

\newcommand{\kB}{k_\text{\tiny B}}
\newcommand{\xb}{{\bf x}}
\newcommand{\adag}{a^\dag}
\newcommand{\adaga}{a^\dag a}

\newcommand\D{\operatorname{d}\!}

\begin{document}

\title{From basic science to technological development: the case for two avenues}
\author{Matteo Carlesso and Mauro Paternostro   \\  \normalsize{Centre for Quantum Materials and Technologies,}    \\
\normalsize{School of Mathematics and Physics, Queens University,}\\ \normalsize{ Belfast BT7 1NN, United Kingdom}}
\date{\today}

\maketitle

\begin{abstract}
We argue that, in the quest for the translation of fundamental research into actual quantum technologies, two avenues that have -- so far -- only partly explored should be pursued vigorously. On first entails that the study of energetics at the fundamental quantum level holds the promises for the design of a generation of more energy-efficient quantum devices. On second route to pursue implies a more structural hybridisation of quantum dynamics with data science techniques and tools, for a more powerful framework for quantum information processing. 
\end{abstract}

\maketitle

\section*{Introduction}

{The promises of quantum technologies seem limitless: from ultra-secure communication to more accurate weather prediction~\cite{weather}, from the design of target-specific drug~\cite{drug,drug2} to  new and revolutionary materials~\cite{materials}. The ambitions for quantum technologies encompass innovations spread broadly across quantum information science, materials, and sensing technologies.}

{There is little doubt that the development of a successful framework for quantum technologies {\it should} build on the very substantial body of work that has been produced, over the course of the past thirty years, on basic science. The best approach is always to put science first. Areas of particular significance in this context include quantum information theory, the foundations of quantum mechanics, material science, computer engineering, and the framework of system identification and process characterisation. However, it also includes research that goes beyond the field of pure quantum technologies, such as the study of quantum effects in thermodynamic or biological processes. While the community has identified four main strands for research and development that will enable the construction of a comprehensive work infrastructure for quantum technologies, the gain of new knowledge underpinning the activities of such {\it pillar}, and thus contributing to technological advances and new applications in the long term, has to stem from basic science.}

{Solving basic science problems in quantum science across all its complexities will enable transformative scientific and industrial progress that, in time, will transition into major drivers for scientific advancement, economy, and even national security.}

{The provision of a comprehensive overview of the various directions along which a study of fundamental physics could lead to technological advancement goes well beyond the scopes of this work. Instead, we shall aim at arguing in favour of two specific areas of investigation, thermodynamics of quantum processes and machine learning for quantum problems, which have recently gained considerable attention from the community interested in the control of quantum systems and dynamics, and comment on how they are significantly contributing to the process of {\it virtuous} two-way influence between fundamental science and technological developments.}\\

{The thermodynamic interpretation of fundamental energy-exchange process occurring among the elements of a quantum device allows for the establishment of fundamental constraint to the efficiency of certain information-processing tasks, most remarkably quantum computation.  This will contribute to the technological step-change entailed by the Second Quantum Revolution by addressing the energetic and entropic footprints of quantum devices. It thus addresses a blind-spot in the work program that followed the publication of the “Quantum Manifesto” upon which the EU-funded Quantum Flagship initiative has been built that must be addressed to achieve the paradigm shift promised by quantum technologies, and thus bypass the energy-consumption issues in information technologies such as those based on CMOS technology. In doing so, the development of a thermodynamic approach to the energetics of quantum information processing will allow:
\begin{itemize}
\item[(1)] The assessment of the energy cost of creating and consuming quantum coherence in the experimental platforms that embody the building blocks of upcoming quantum processors. Their thermodynamic assessment will allow for the quantification of the energetic cost of processing quantum information, including computational tasks; 
\item[(2)] The establishment of a framework for the exploration and understanding of the interplay between logical and thermodynamic irreversibility, which is the fundamental source of heat dissipation in computing systems.  
\end{itemize}
}

{From a different, yet related perspective, the interplay between data-intensive research and {\it quantum} is leading to the development of new hybrid methods for control and performance-optimisation of quantum processes, while opening up the possibility to design less expensive methods for system identification, performance validation and quantum-property reconstruction. The use of artificial intelligence in the management of quantum devices widens the already significant pool of hardware problems benefiting of machine learning, providing a boost to the development of near-term quantum devices enhanced by artificial intelligence accelerators. In time, it will also allow for the use of quantum computing devices as neural networks for the systematic adaptation of the controls of quantum processors, the classification of datasets resulting from quantum processes, and the inexpensive reconstruction of states or properties of quantum devices.}\\

\section*{Thermodynamics for an energetically efficient quantum information processing}

The application of quantum mechanics to the mesoscopic and macroscopic scales faces the limitations imposed by the surrounding environment, which can have a strong influence on the quantum system that one wants to study or employ in quantum technology applications. 
The comprehension of the interaction of such an open quantum system with its environment becomes pivotal for the development and  enhancement of quantum technologies as quantum sensing and communication.\\

To be eventually able to suppress the action of the environment in open quantum systems, one needs to identify all -- or at least the main --  contributions. Clearly, depending on the quantum system under scrutiny, some environmental mechanisms will contribute more than other to the total influence.  These effects can be studied with the mathematical tools provided by the open quantum system theory. In such a framework, the system S is considered in interaction with a surrounding environment E with a total evolution defined by a unitary dynamics $\hat U_t$ with respect to the total Hamiltonian of the system + environment. Namely, the total statistical operator evolves as $\hat \rho_\text{\tiny SE}(t)=\hat U_t\hat \rho_\text{\tiny SE}(0)\hat U_t^\dag$. Then, one averages over the degrees of freedom of the environment to get the effective (reduced) dynamics of the system alone. This is usually described by the dynamical equation known as  master equation $\frac{\D \hat \rho_\text{\tiny S}(t)}{\D t}=\mathbb L_t[\hat \rho_\text{\tiny S}]$, whose form is defined by the generator $\mathbb L_t$. The latter can encode decoherence (loss of quantum coherence and decay of quantum superpositions of the system, namely suppresses the quantum features of the system) and  dissipation (energy exchange between system and environment)   due to the environment. Typically, the decoherence is described through the master equation (in position representation)~\cite{JZ}
\begin{equation}
\frac{\D \braket{x|\hat \rho_\text{\tiny S}(t)|x'}}{\D t}=-\tfrac i \hbar \braket{x|[\hat H,\hat \rho_\text{\tiny S}(t)]|x'}-\Gamma(x-x')\braket{x|\hat \rho_\text{\tiny S}(t)|x'}, 
\end{equation}
where the rate $\Gamma(x-x')$ quantifies the deviations from the unitary dynamics defined by the system's Hamiltonian $\hat H$. 
\begin{figure}
\centering
\includegraphics[width=0.7\columnwidth]{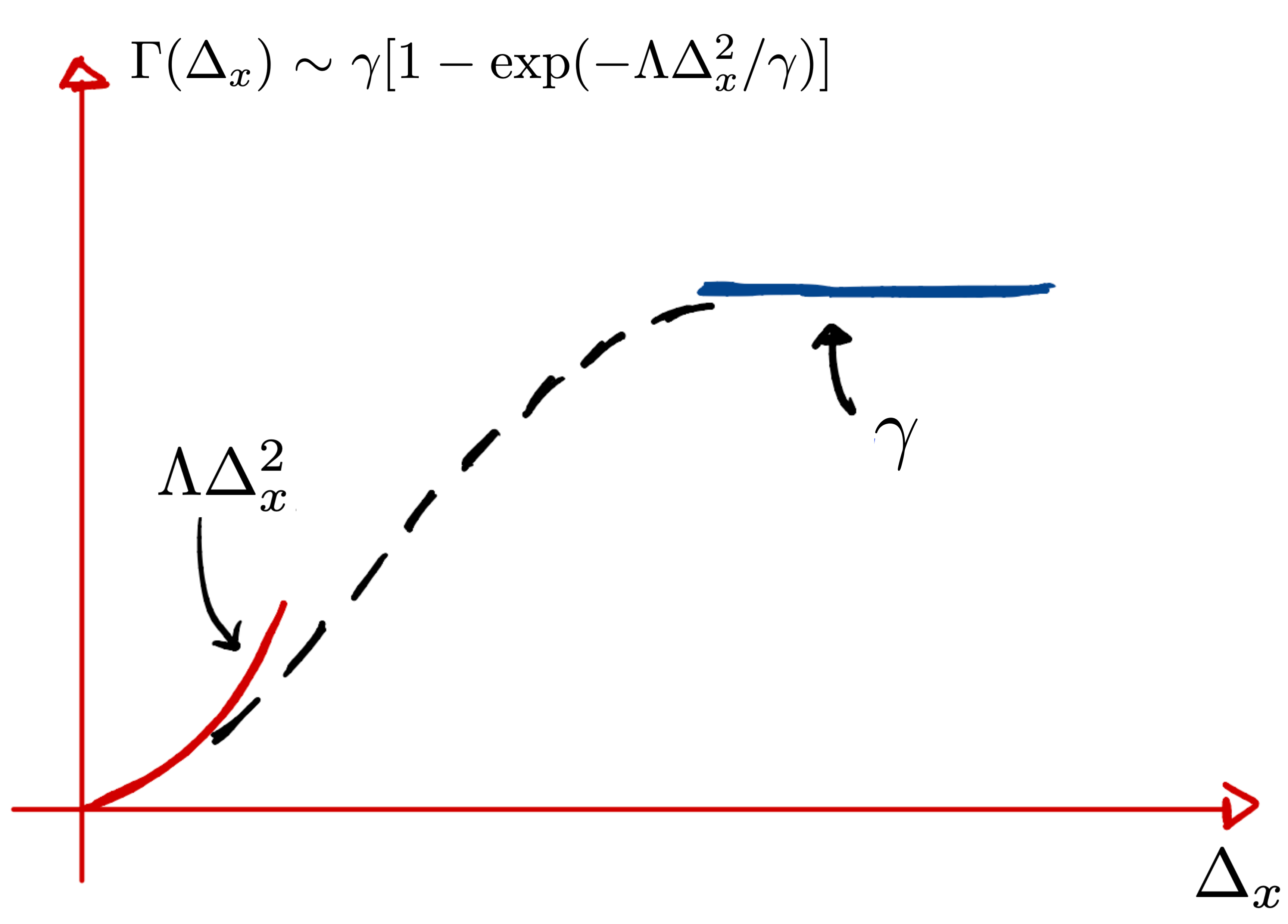}
\caption{\label{fig.behaviourGamma} The decoherence function $\Gamma$ as a function of the superposition distance $\Delta_x$ in the two relevant limits. In the long-wavelength limit $\Gamma(\Delta_x)\sim\Lambda\Delta_x^2$ (red line), while in the short-wavelength limit $\Gamma(\Delta_x)\sim \gamma$ (blue line). The dashed grey line represent a possible parametrisation of $\Gamma(\Delta_x)$ that connects the two limiting cases. 
\cite{noi2}.
}
\end{figure}
\begin{figure}
\centering
\includegraphics[width=0.8\columnwidth]{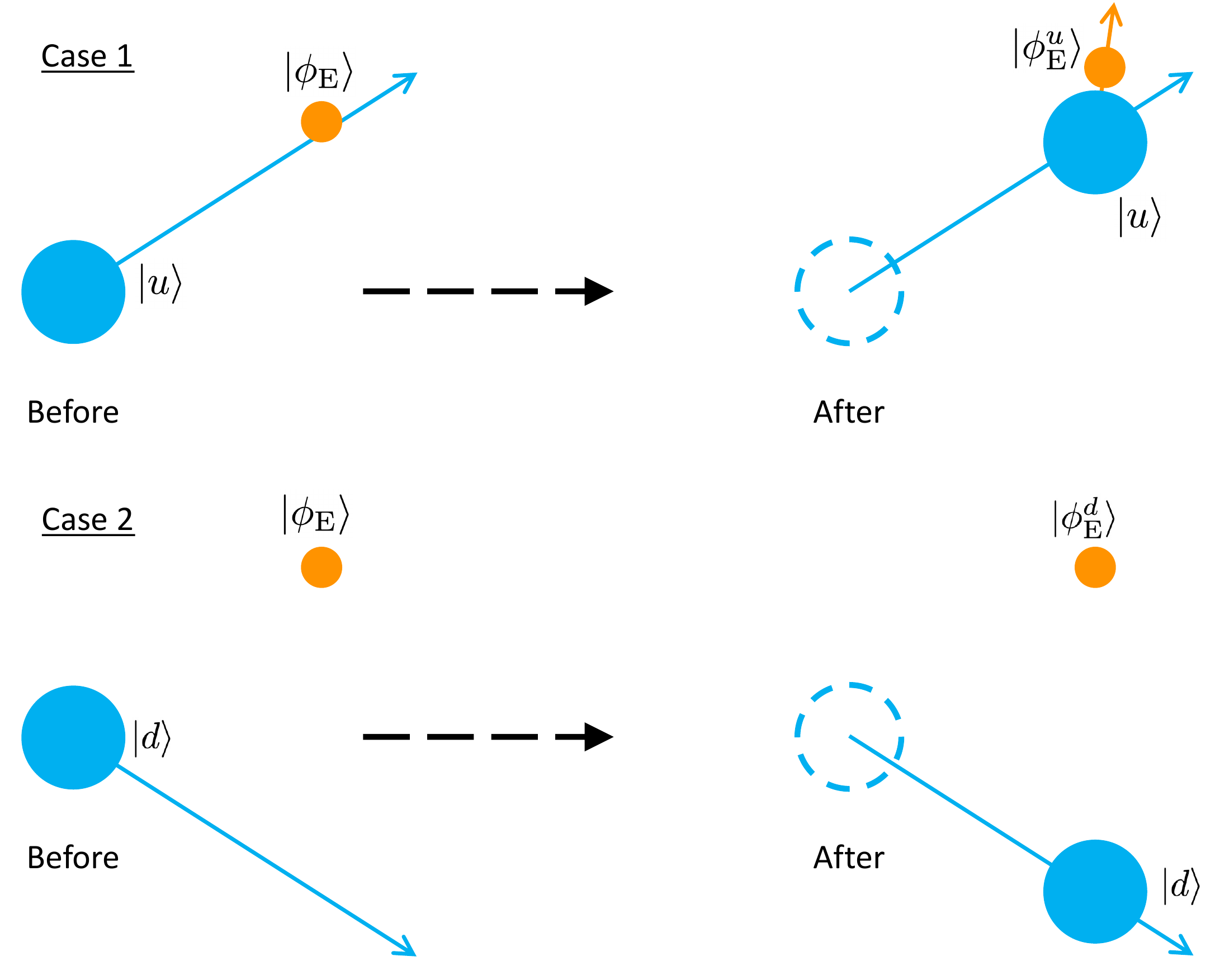}
\caption{\label{fig.collision} Scheme of the collisional decoherence. Case 1: the system is initially prepared in the state $\ket u$, then it will collide with the gas particle which is in the state $\ket{\phi_\text{E}}$, eventually ending in the state $\ket{\phi_\text{E}^u}$. Case 2: conversely, if the system is initially prepared in $\ket d$, it will leave the state $\ket{\phi_\text{E}}=\ket{\phi_\text{E}^d}$ untouched.}
\end{figure}
Such a rate $\Gamma$ scales quadratically with $\Delta_x=|x-x'|$ for small spatial separations with respect to the characteristic length-scale of the specific decoherence process (namely, the long wavelength limit), while it saturates to a constant value for large separations (short wavelength limit). A way to connect these two limits is to approximate the decoherence rate to $\Gamma(\Delta_x)\sim\gamma[1-\exp(-\Lambda \Delta_x^2/\gamma)]$, which introduces the rate of events $\gamma$ and the diffusion constant $\Lambda$ [cf. Fig.~\ref{fig.behaviourGamma}].
The typical example of such a separation of length-scales is provided by the collisional decoherence~\cite{Schlossauer}.
This is the case  of a mesoscopic particle, having the role of the system of interest, under the action of the collisions of the particles constituting the residual gas in the chamber where the experiment is performed. Let us assume that the initial state of the system $\ket\psi$ is in a superposition of the particle going up $\ket{{u}}$ and going down $\ket{{d}}$: $\ket \psi\propto\ket{u}+\ket{d}$, and that the single collision with the gas particle takes place only if the systems is in the state $\ket{u}$ (see Fig.~\ref{fig.collision}). In particular, one can  describe the collision as $\ket{u}\otimes\ket{\phi_\text{E}}\to\ket{u}\otimes\ket{\phi^u_\text{E}}$.
\begin{table}[b!]
\caption{Decoherence rates $\Gamma(\Delta_x)$ in the long ($\Gamma(\Delta_x)\sim\Lambda\Delta_x^2$) and short ($\Gamma(\Delta_x)\sim\gamma$) wavelength limits with respect to the superposition distance $\Delta_x$ \cite{PhysRevA.84.052121}. The collisions with the residual gas are characterized by the length scale $2\pi \hbar/\sqrt{2\pi m_\text{p}\kB T}$, while scattering, absorption and emission of blackbody radiation by $\pi^{2/3}\hbar c/\kB T$. Here, we distinguish the internal temperature $T_\text{i}$ (relevant for the emission process) from the external temperature $T_\text{e}$ (relevant for the absorption process).}\label{wrap-tab:1}
\begin{tabular}{c|c|c}
\hline\hline
\ &$\Lambda$ (long wavelength limit)&$\gamma$ (short wavelength limit) \\
\hline
Collisions&$\frac{8\sqrt{2\pi}m_\text{p} \bar v P R^2}{3\sqrt{3}\hbar^2}$&$\frac{16\pi\sqrt{2\pi}PR^2}{\sqrt{3}\bar v m_\text{p}} $\\  
Scattering&$8!\frac{8 \zeta(9)}{9\pi}R^6 c\,\Re\left(\frac{\epsilon-1}{\epsilon+2}\right)^2\left(\frac{\kB T}{\hbar c}\right)^9$&$8!\frac{8 \zeta(9)\pi^{1/3}}{9}R^6 c\,\Re\left(\frac{\epsilon-1}{\epsilon+2}\right)^2\left(\frac{\kB T}{\hbar c}\right)^7$\\  
Abs.\&Emis.&$\frac{16\pi^5}{189}R^3 c\,\Im\left(\frac{\epsilon-1}{\epsilon+2}\right)\left(\frac{\kB T_\text{(i,e)}}{\hbar c}\right)^6$&$\frac{16\pi^6\pi^{1/3}}{189}R^3 c\,\Im\left(\frac{\epsilon-1}{\epsilon+2}\right)\left(\frac{\kB T_\text{(i,e)}}{\hbar c}\right)^4$\\
\hline\hline
\end{tabular}
\end{table} 
Now, the collision on the full superposition reads $\left(\ket{u}+\ket{d}\right)\otimes\ket{\phi_\text{E}}\to\ket{u}\otimes\ket{\phi^u_\text{E}}+\ket{d}\otimes\ket{\phi^d_\text{E}}$, where $\ket{\phi^d_\text{E}}=\ket{\phi_\text{E}}$. Correspondingly, the reduced density matrix (obtained after averaging over the environmental degrees of freedom) in the $\{\ket u,\ket d\}$ basis will have off-diagonal terms being proportional to $\braket{\phi_\text{E}^u|\phi_\text{E}^d}$. Now, it is clear that more the states $\ket{\phi_\text{E}^u}$ and $\ket{\phi_\text{E}^d}$ differ each other stronger becomes the decoherence effect, which is eventually maximised when the two environmental states are orthogonal. The difference between such states is imposed by the momentum transfer between the system and the gas particle during the collision, and here comes into play the ratio between the characteristic length-scale of the gas particle $ 2\pi \hbar/\sqrt{2\pi m_\text{p}\kB T}$ (where $\hbar $ is the reduced Planck constant, $m_\text{p}$ is the mass of the gas particle, $\kB$ is the Boltzmann constant and $T$ is the temperature of the environment).
Correspondingly, the decoherence effects scale with $m_\text{p}\bar v P R^2\Delta_x^2$ in the long wavelength limit (depending on the pressure $P$ in the experiment, the mean velocity $\bar v$ of the gas and the linear dimension $2R$ of the system). 
 Conversely, in the short wavelength limit go as $P R^2/\bar v m_\text{p}$. The effects of collisional decoherence are particularly important in matter-wave interferometry experiments~\cite{hornberger}, which have been key to establish the emergence of quantum effects at the mesoscopic scale by dealing with system of increasing size and complexity. \\

Another example of decoherence is provided by three processes involving blackbody radiation, which are scattering, absorption and emission, and  are characterized by the  length-scale $\pi^{2/3}\hbar c/\kB T$.
The first among these three processes is the equivalent of the collisions with the residual gas particle: the blackbody radiation elastically scatters on the system and there is a momentum exchange between the two. Conversely, during the absorption, the scattering is completely inelastic, the photon is absorbed and the system gains completely the photon's momentum. Finally, the emission process can be understood as the exactly the time-inverted process of the absorption, where a photon is emitted and the particle losses momentum.
Depending on the process, they are characterized by a different scaling in the long wavelength limit: $R^6 T^9 \Re(\epsilon-1/\epsilon+2)^2$ for the scattering and $R^3 T^6 \Im(\epsilon-1/\epsilon+2)$ for emission and absorption where $\epsilon$ is the dielectric constant of the system. Similarly, in the short wavelength limit, one has $R^6 T^7 \Re(\epsilon-1/\epsilon+2)$ for the scattering and $R^3 T^4 \Im(\epsilon-1/\epsilon+2)$ for emission and absorption. \\

When dealing with much smaller systems, where the above mentioned decoherence mechanisms can be negligible, other mechanisms kick in. For example, in Bose-Einstein condensates (BEC), the main concern in loosing quantum properties is the reduction of the number of particles in the condensate rather than decoherence understood as suppression of quantum coherences. Here, two processes are relevant. The first concerns the collisions of the atoms in the condensates with those of the background thermal could, which is made of the atoms of the cloud that do not appertain to the condensate. The second process is that of the three-body recombination, which accounts for the inelastic collisions between atoms in the BEC that lead to the molecule formation. These two processes can be described together  through \cite{laburthe2003observation}
\begin{equation}
    \frac{\D N_t}{\D t}=-K_1 N_t-\tilde{K}N_t^3,
\end{equation}
where $K_1$ is the one-body decay rate due to the interaction with the thermal cloud and $\tilde{K}=K_3/((2\pi)^3 3^{3/2} \sigma_t^6)$ depends on the position variance $\sigma_t$ of the BEC and the three-body loss coefficient $K_3$. For large BECs, the stronger reduction effect comes from the three-body recombination process, thus indicating that to generate and maintain a large BEC for long times becomes pivotal to reduce $\tilde K$ as much as possible. To be quantitative, for keeping the reduction below the 10\% over 10\,s when assuming $K_1\sim10^{-3}\,$s$^{-1}$, one requires $\tilde K\sim10^{-12}\,$s$^{-1}$ for an initial value of atoms of $N_0\sim10^5$, while $\tilde K$ needs to drop below $4\times10^{-14}\,$s$^{-1}$ when considering $N_0\sim 5\times 10^{5}$. This corresponds to almost two orders of magnitude reductions of $\tilde K$ when one considers \textit{only} an increase of a factor five in the number of atoms in the condensate. 
\\

The progress of open quantum system's theory has provided the grounds to quantify the limiting effects imposed by the  environment surrounding a system, and thus carefully characterize the way the former might affect quantum technologies. Typically, the process of interaction between a quantum system and its environment implies the exchange of energy among the parties. The washing out of the degrees of freedom of the environment results in such process to be interpreted as an exchange of heat, which might eventually lead it to equilibration and, in the case of an environment already at thermal equilibrium, thermalisation of the system. \\

\begin{figure}
\centering
\includegraphics[width=0.8\columnwidth]{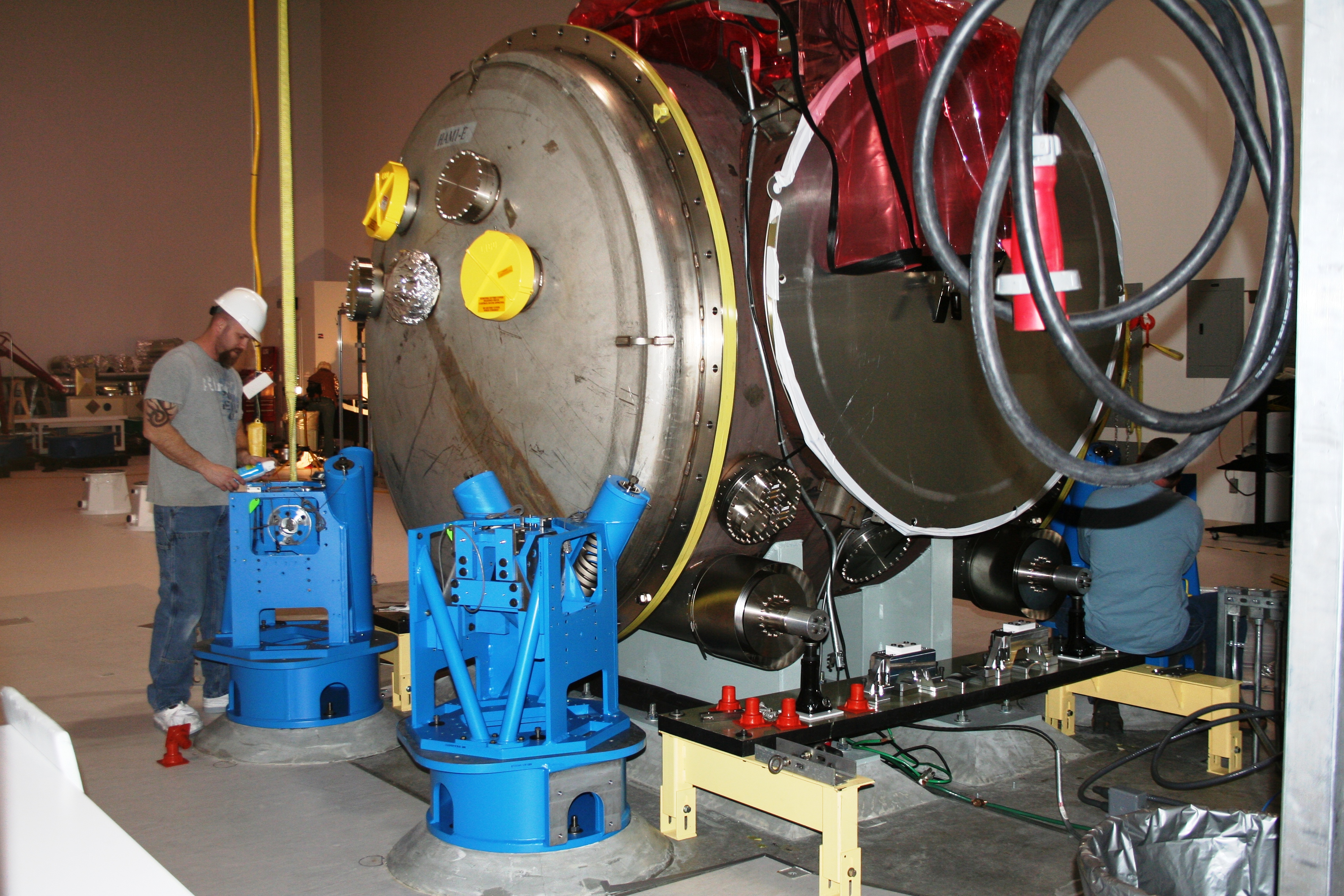}
\includegraphics[width=0.8\columnwidth]{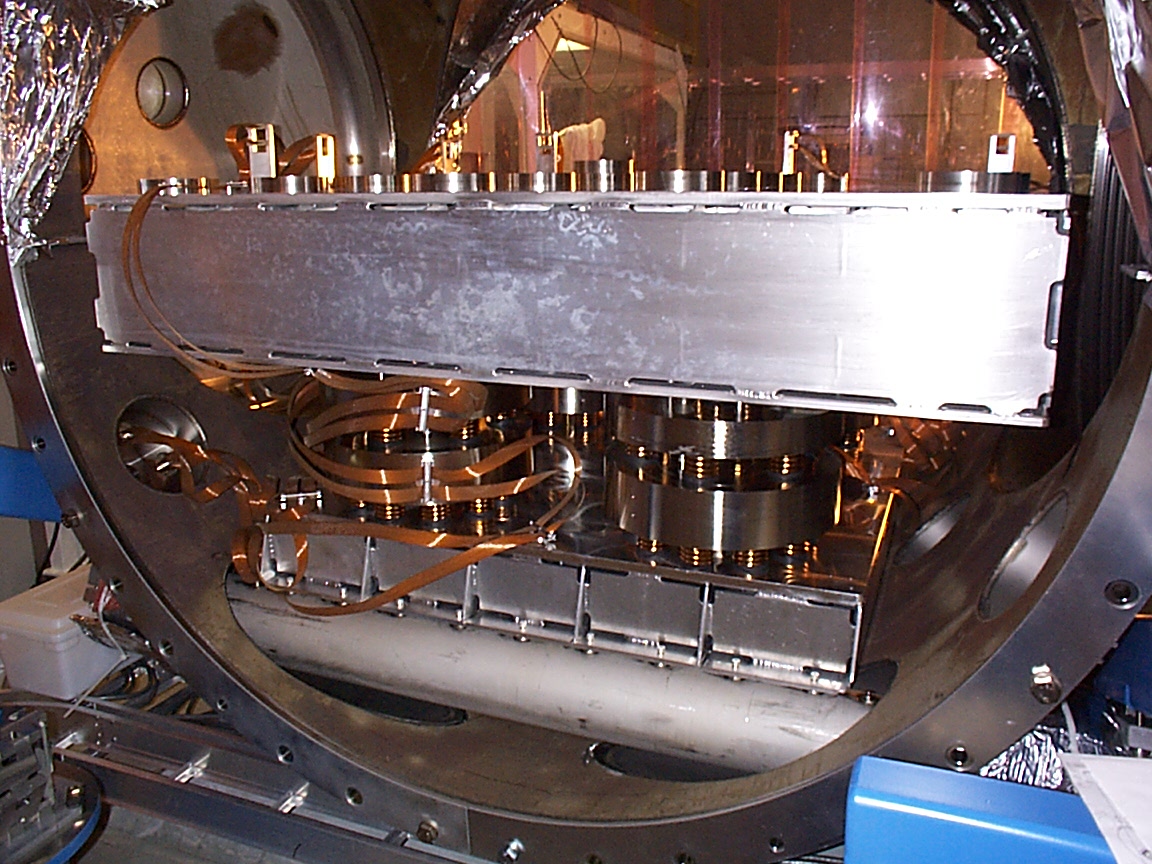}
\caption{\label{fig.springs} Suspensions of the Advanced LIGO's mirrors. Top figure: the Hydraulic External Pre-Isolator (HEPI) in blue provides the first stage of isolation outside the vacuum chamber where the LIGO's mirror are placed. Bottom figure: 
the Seismic Attenuation System (SAS) within the vacuum chamber constitutes the second seismic isolation.
 Figures from  \cite{LIGOweb}.}
\end{figure}
Such an interaction moves the state of a system away from those delicate conditions necessary for the emergence of fragile genuine quantum effects. Usually, one tries to depart from such a hindering situation and keep the system as  isolated as possible. This implies lowering the temperature of the system thus avoiding high-temperature thermal states, reducing the pressure of the surrounding residual gas -- thus minimizing their collisions with the system -- and isolate the system from all possible noises that could perturb it. Other than the noises from the surrounding environment meant as residual gas and blackbody radiation, one needs to account in the total noise budget also those imposed by the specific experimental setup that is used. Among the different sources, we can mention the vibrational noises that can have various sources as the Earth seismic activity that is particularly relevant at low frequencies, the urban traffic and the refrigerator that can be predominant at the corresponding characteristic frequencies.  Some experiments are so sensitive that can feel variations of the local gravitational field when, for example, a train passes through the nearby train station. 
To mitigate the vibrational noise, one typically detaches as much as possible the system from ground by employing a system of springs that attenuate such noises at the relevant frequencies. Figure \ref{fig.springs} shows a possible implementation of these spring systems when applied to the mirrors of the gravitational wave detector LIGO  \cite{LIGOweb}. Alternatively, one can perform experiments in free-fall where several components of vibrational noises are avoided completely. One option is constituted by experiments performed in drop towers as that of ZARM in Bremen \cite{ZARM} or the Einstein Elevator \cite{EinsteinElevator}, which can provide up to around 9\,s and 4\,s of free-fall respectively. However, this option undergoes to strong limitations in the time for the run of the experiment, which is determined by the height of the drop tower, and it would be still subject to vibrations due to the friction between the residual gas in the drop tower and the capsule containing the actual experiment. Another option is to perform experiments in space \cite{CommentNature, noi2, noi, MAQRO, QcomSpace}, which would provide much longer free-evolution times and it would avoid the problem of the friction. To be quantitative, LISA Pathfinder --- which is the prototipe for the planned space-based gravitational wave detector LISA --- has demonstrated an acceleration noise floor of $10^{-15}\,$m\,s$^{-2}/\sqrt{\text{Hz}}$ that constitutes an improvement of around ten orders of magnitude with respect to the value of $10^{-5}\,$m\,s$^{-2}/\sqrt{\text{Hz}}$ provided by the ZARM drop tower
 \cite{Selig}.\\

Being able to control the environment is pivotal to possible quantum technological applications such as quantum sensing or quantum communication, but it is also the basis to perform test of basic science as the detection of gravitational waves or the generation of quantum superposition with massive objects. Science-driven experiments demand and, with time, generate a significant progress, at both the fundamental and technological level, in the design and the achievement of the  conditions able to meet the requirements for a very low-noise, controlled operational regime~\cite{noi,noi2}. Examples are provided by the thirty-years endeavour of gravitational wave detector such as LIGO (Laser Interferometer Gravitational-wave Observatory) or --- more recently --- by its space counterpart LISA (Laser Interferometer Space Antenna) \cite{LISA} that is planned to be launched in 2037. Here, the noise's reduction is fundamental to being able to detect the fainted signals of the passage of gravitational waves. Other science-driven experiments are  focused on the detection of non-standard effects appearing as noises conversely to a signal as in the case of gravitational waves. In such a case, the challenge is to distinguish one (non-standard) noise within a background of other noises. Examples of such effects run from deviations of quantum mechanics \cite{revcollasso,revcollasso2}, gravitational decoherence \cite{gravdecoherence}, quantum gravity effects \cite{quantumgravreview},
dark matter/energy \cite{PhysRevLett.125.181102} and others.
\\

\begin{figure}
\centering
\subfigure[resolved sideband cooling]
{\includegraphics[width=0.8\columnwidth]{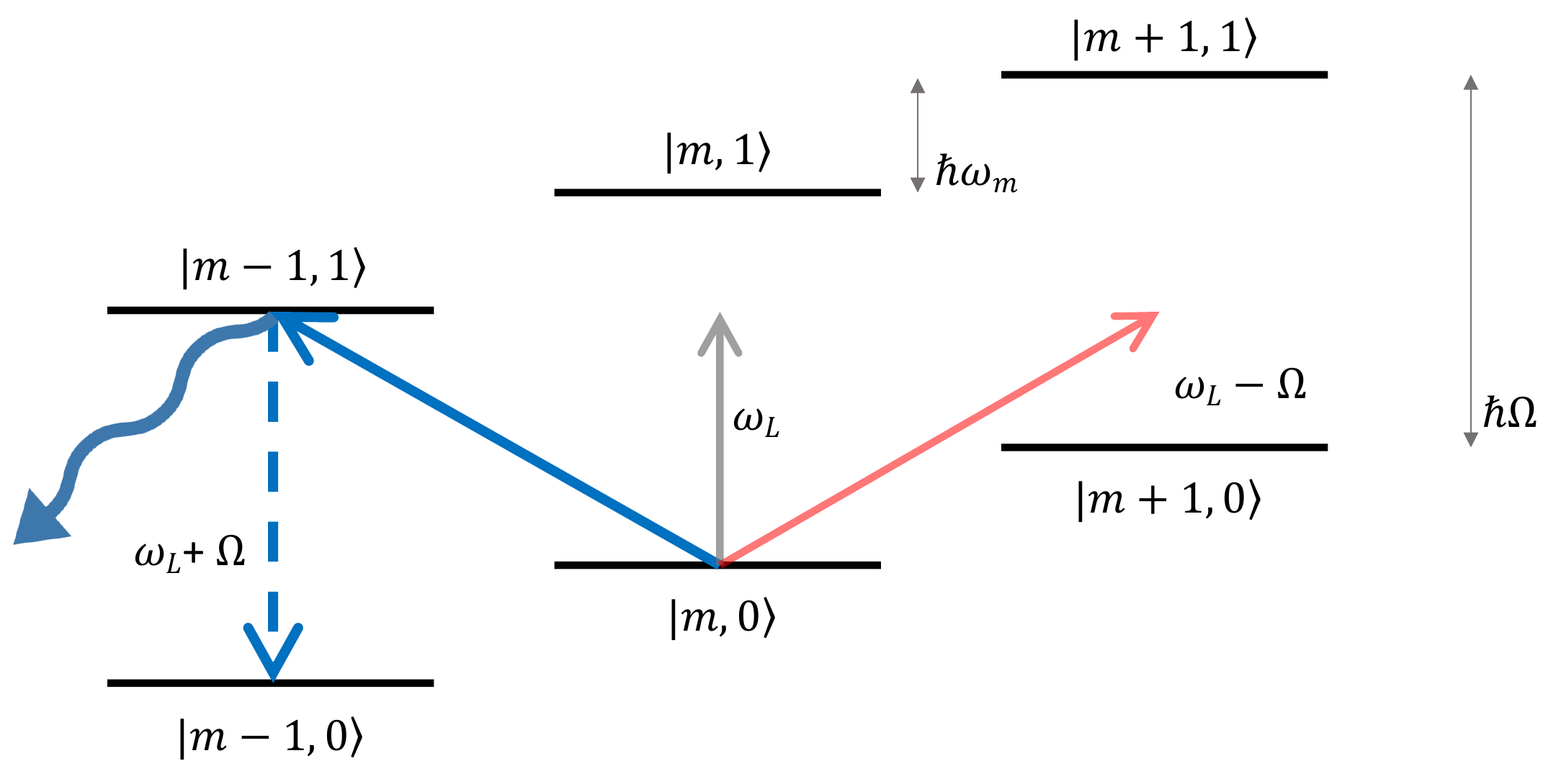}
}
\hrule
\subfigure[parametric feedback cooling]{
\includegraphics[width=0.7\columnwidth]{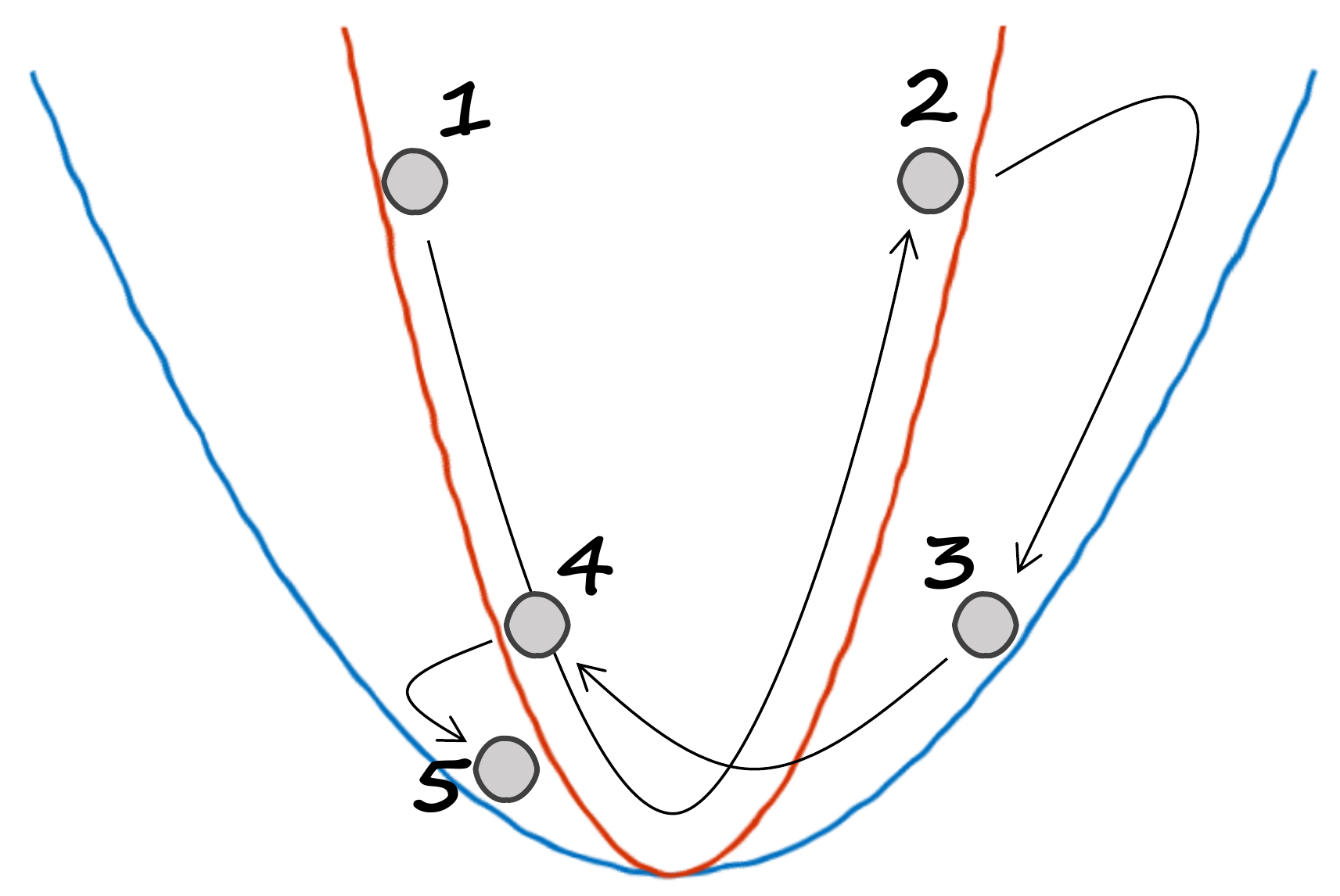}
}
\caption{\label{fig.cooling} Conceptual scheme of the working process behind the resolved sideband cooling (a) and the parametric feedback cooling (b).}
\end{figure}
Clearly, in some situations, changing the environmental parameters, such as temperature and pressure, is not sufficient to achieve the desired conditions. The paradigmatic example is the achievement of the ground state in optomechanical systems~\cite{NatPhys}, which is something hardly possible by simply reducing the temperature of the environment, and one needs to  find alternative routes toward this aim. Indeed, satisfying the condition of $\hbar\omega_\text{m}\gg\kB T$, where $\omega_\text{m}$ is the mechanical frequency of the system harmonically trapped, is challenging also for dilution refrigerators providing a 20\,mK environment as one would require $\omega_\text{m}$ to be larger than 10\,GHz. Active cooling techniques were thus developed, such as, among others, resolved sideband or feedback cooling in optomechanical system.

In the case of sideband cooling (top panel of Figure \ref{fig.cooling}), the mechanical resonator, which plays the role of the main system to be cooled and characterized by the frequency $\omega_\text{m}$, is placed in mutual interaction with an optical field of frequency $\Omega$. Both the systems are coupled to the same thermal bath at temperature $T$. However, since the frequency of the optical field $\Omega$ is assumed to be much larger than that of the resonator $\omega_\text{m}$ and such that $\hbar\Omega\gg\kB T$, from the perspective of the optical field, the bath is in the vacuum state, i.e. in the ground state. This means that the optical field can only lose energy, at rate $\kappa$ towards the bath without gaining any. Then, by employing an external laser at frequency $\omega_\text{L}$, one can stimulate the transition between the $|m,0\rangle$ state, which indicates $m$ excitations in the mechanical resonator and $0$ in the optical field, and the $|m-1,1\rangle$; then, in the limit of $\kappa\gg\omega_\text{m}$, the optical field will decay from $|1\rangle$ to $|0\rangle$, emitting the photon in the thermal bath, before the excitation goes back to the mechanical resonator. To summarise, the process leads to the following transition
\begin{equation}
    |m,0\rangle\to|m-1,1\rangle\to|m-1,0\rangle,
\end{equation}
meaning that effectively the mechanical resonator has lost an excitation. If the rate of this transition process, being $g^2/(\kappa/2)$ where $g$ is the coupling between the mechanical resonator and the optical field, is larger than the effective heating $\gamma\bar n(\omega_\text{m})/2$ of the mechanical resonator, where $\bar n$ is the thermal occupation number, one can cool the system's occupation number to the value of
\begin{equation}
    n=\bar n(\omega_\text{m})\frac{\kappa\gamma}{4g^2}+\left(\frac{\kappa}{4\omega_\text{m}}\right)^2,
\end{equation}
which can go below the unity, indicating that the ground state cooling regime has been reached.

The concept behind the feedback cooling is instead completely different. In such a case, one measures one among the observables of the system -- for example a quadrature or the energy -- and it acts on it depending on the measurement outcome. In particular, the feedback process can be implemented directly through the use of the measurement apparatus, since it already acts on the system inducing backaction effects. Such effects can be however modified to cool the system towards the ground state. An example is given by the case of an harmonic oscillator whose momentum is continuously monitored \cite{Vovrosh}. The outcome of such a position $\langle \hat x\rangle$ is then used in the feedback loop to generate the feedback force, which will act on the harmonic oscillator and it is implemented by modifying the stiffness of the harmonic trap. In particular, such a force  is proportional to the measurement outcome \begin{equation}
    F_\text{bf}= -k_\text{fb}(t)\langle\hat x\rangle,
\end{equation}
where $k_\text{bf}(t)$ is the change of the stiffness due to the feedback. To reduce the amplitude of motion of the particle, that we assume can be parametrised harmonically as $x(t)=\langle \hat x(t)\rangle=x_0 \sin(\omega_\text{m} t)$, the trap stiffness is increased as the particle climbs the potential well so that  its kinetical energy is reduced. When instead the particle goes towards the center, the trap stiffness is reduced. In particular, one needs to modify the stiffness at twice the frequency of the harmonic oscillator to achieve the energy reduction, which leads to an effective extra damping $\delta\Gamma$ with respect to that of the case without feedback $\Gamma$. Eventually, one can reach the center-of-mass temperature 
\begin{equation}
    T_\text{c.m.}=\frac{\Gamma}{\Gamma+\delta\Gamma}\, T,
\end{equation} 
which is lower than the environmental temperature $T$.\\

\begin{figure}
\centering
\subfigure[Evaporative cooling]
{\includegraphics[width=0.9\columnwidth]{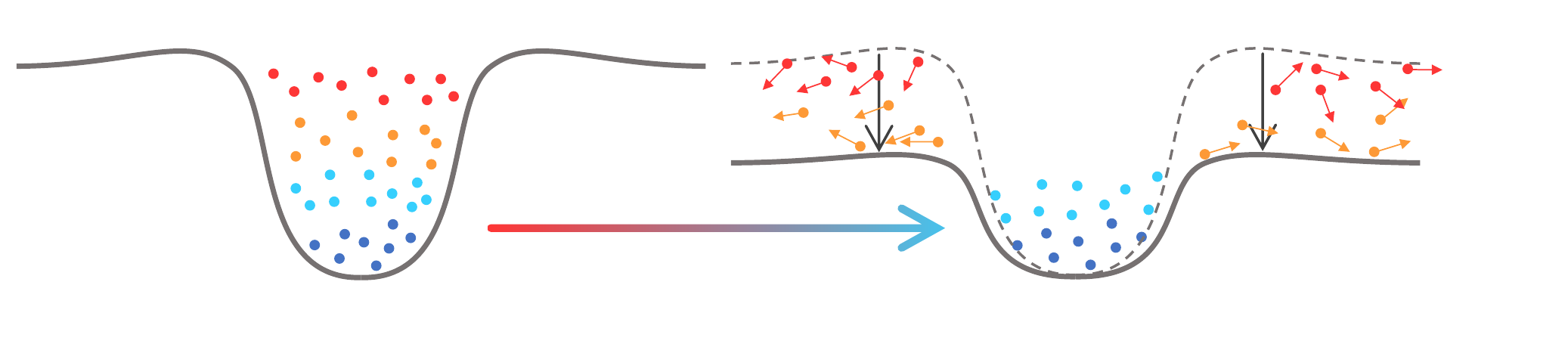}
}
\hrule
\subfigure[Doppler cooling]
{\includegraphics[width=0.9\columnwidth]{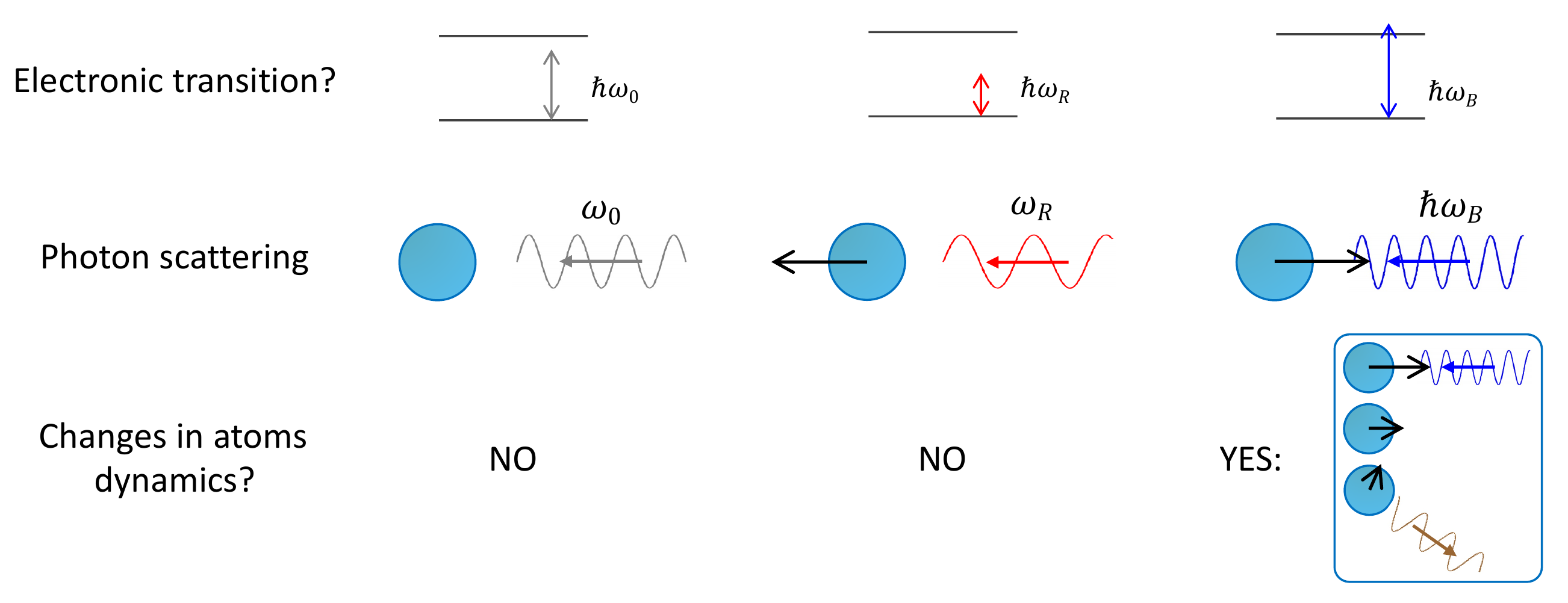}
}
\hrule
\subfigure[Delta-kick cooling]
{\includegraphics[width=0.9\columnwidth]{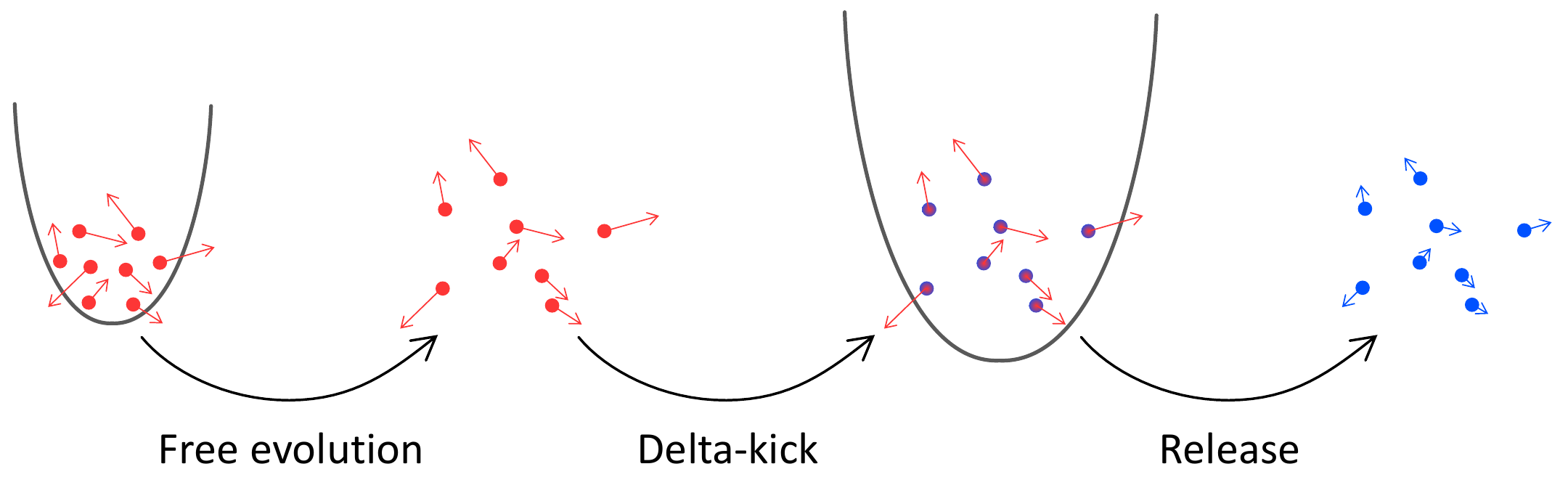}
}
\caption{\label{fig.variouscooling} Conceptual scheme of the working process behind the evaporative cooling (a), Doppler cooling (b), and delta-kick cooling (c).}
\end{figure}
Cooling techniques were developed also for other quantum systems as for Bose-Einstein condensates (BEC) as, for example, the evaporative cooling, the Doppler laser cooling or the delta-kick cooling (see Fig.~\ref{fig.variouscooling}].
In the case of the evaporative cooling \cite{Petrich}, the atomic cloud is initially trapped and let thermalise. Then, the potential that traps the cloud is ramped down so that the more energetic atoms can escape the trap boundaries. Those atoms that remain in the trap are those with less kinetic energy, meaning that the corresponding energy distribution can be associated to a lower temperature.

The second cooling techniques for atomic systems is the Doppler laser cooling, which is based on the Doppler effect \cite{doppler1,doppler2}. Here, one shines the atoms with a laser having a frequency slightly below the electron transition, namely it is a red detuned laser. This means that the electrons of a laser will not absorb the corresponding photon, as its energy is not sufficient to lead to an energetic transition. Then, one here exploits the Doppler effect: the atom moving away from the laser laser source will see the photons being even more red detuned (i.e. at lower frequency), while if it moves toward the source it will see it blue detuned (i.e. at higher frequency). Then, when moving away, the atom's electron will not get excited; on the contrary, when moving towards the laser source the gap between the transition energy and the initially red detuned laser is filled and they can absorb the photon. Due to the conservation of momentum, the atom will then slower and, eventually, its electron will spontaneously emit the absorbed photon. Since the direction of the emission will be random, there will be no net increase of energy when taking the average over the ensemble of atoms.

The third cooling technique we mention is the delta-kick cooling \cite{Ammann}. Here, one neatly traps the ensemble of atoms and let them thermalize. Then, one leaves the atoms to evolve freely for some limited time before switching on a more broad harmonic trap. During the free evolution, the atoms having more kinetic energy will move further away from the center of the trap. This is then exploited by the harmonic trap, as it will imprint a force on each atom that is proportional to the distance just covered $F\propto-x$. Therefore, more energetic atoms will get a stronger ``kick'' back toward the center of the trap, thus strongly cooling the entire atom cloud.
\\

Despite the efforts along the lines illustrated above, and the enormous progress achieved in the construction of devices that are only mildly affected by environmental effects, any piece of quantum technology will have to be considered as intrinsically and inherently noisy. As such, it is only meaningful to address the challenge posed by the coupling between a quantum system and its environment from a thermodynamic perspective that treats the energetics of quantum processes from a fundamental perspective but aims directly to tangible technological development~\cite{book}.

Thermodynamics was developed in the 19th Century to improve the efficiency of steam engines. Its impact has been enormous, and affected fundamental science, technology and everyday life alike. In the 3rd Millennium, the challenge is to address quantum thermodynamics to design efficient quantum machines. Quantum thermodynamics (QT) describes stochastic processes of a quantum system out of its state of equilibrium. The intrinsic stochastic nature of the process can be described in terms of probability density functions, which are characterized by quantum fluctuation relations. These regulate the amount of work and heat that can be respectively performed or exchange by the system in an arbitrary non-equilibrium process. Such an amount is bounded by the thermodynamic irreversibility of the process (equivalently to the second principle of thermodynamics) and can be characterized in terms of the entropy production (EP) and the entropy production rate (EPR)~\cite{RMP}. Under this perspective, QT provides the means to identify among different protocols for state preparation which is the most efficient. Indeed, it has been shown that, in a heat engine, the efficiency of a cycle is given by the Carnot efficiency reduced by a term proportional to the EPR~\cite{RMP}. Thus, at minimum EPR corresponds maximum efficiency, and this can be exploited to determine the most efficient protocol.
The irreversibility characterisation provided by quantum thermodynamics can be used to quantify the performance cost of quantum protocols, and thus find the optimal one with respect to suitable figures of merit, as for example the energy required to perform a specific protocol.\\

The community working in quantum technology is progressively developing awareness of this challenge~\cite{varia}, although a workplan designed to establish the necessary knowledge-baseline for advancing an energetics-based approach to the design and realisation of quantum devices is still lacking~\cite{auffeves}. Such workplan should be built through the following concrete steps: 
\begin{itemize}
\item[{\it Step 1}]	Design and demonstrate processes that enable the manipulation of energy (work extraction and  heat flux steering)  by means of quantum systems - both elementary and complex ones - against (classical and quantum) noise;
\item[{\it Step 2}]	Identify the best schemes for the enhancement of the performance of such processes through genuine quantum resources (including quantum measurements) and sophisticated quantum control schemes; 
\item[{\it Step 3}]	Benchmark the points above in noisy intermediate scale quantum systems embodying important test-beds for a future generation of larger-scale energy effective devices;
\item[{\it Step 4}] Demonstrate new functionalities for the efficient control of quantal energy-exchange processes, including novel cooling techniques, and strategies to harness heat fluxes resulting from such processes;
\item[{\it Step 5}] The identification of thermodynamics-inspired strategies for the minimisation of the energetic cost of quantum information processing. 
\end{itemize}

In what follows, we will address some of these intended steps in some quantitative details. 

\subsection{On Step 2}

The fundamental connections between information and thermodynamics dates back to the seminal contributions by Maxwell, Szilard, and Landauer. The process of acquiring information can impact the entropic balance of a given physical process. Such information must thus be accounted for and considered on equal footing to other thermodynamic quantities, such as heat and work. This is particularly relevant for processes involving microscopic systems, which are fundamentally dominated by fluctuations: the acquisition of information through measurements introduces additional stochasticity and makes the overall process strongly dependent on the monitoring methodology. When assessing a monitored system, one must distinguish between conditioned and unconditional dynamics, the former being affected by the measurement records~\cite{Zambrini}.

To this end, let us consider the dynamics of a continuously measured open quantum system subjected to a Markovian evolution can be described by a stochastic master equation (SME)
that describes the evolution of the quantum state of the system \textit{conditioned} on the outcomes of the continuous measurement.~\cite{doherty,serafini,genoni}. Upon averaging over all stochastic trajectories, weighted by the outcomes probabilities, the stochastic part vanishes leaving a deterministic Lindblad master equation for the open system dynamics, whose dynamics will be called \textit{unconditional} throughout the paper. 

Solving SMEs is in general a complex goal, made simpler when dealing with Gaussian systems, for which the problem is translated into solving a simpler system of stochastic equations. In order to fix the ideas, let us consider a simple system consisting of a single harmonic oscillator described by a pair of quadrature operators $(\hat{q},\hat{p})$ with $[\hat{q},\hat{p}]=i\hbar$, and define the vector of operators $\hat{\xb}=(\hat{q},\hat{p})$. The generalisation to an $n$-oscillator system has been reported in Ref.~\cite{MP}. In the Gaussian framework, the Hamiltonian is bilinear in the quadratures and can be written as
    $\hat{H}=\frac{1}{2}\hat{\xb}^{T}H_{s}\hat{\xb}+\mathbf{b}^{T}\Omega\hat{\xb}$,
where $H_{s}$ is a $2\times 2$ matrix, $\mathbf{b}$ is a two-dimensional vector accounting for a (possibly time-dependent) linear driving, and $\Omega=i\sigma_{y,j}$ is the symplectic matrix ($\sigma_{y,j}$ is the $y$-Pauli matrix of subsystem $j$).
 By modelling the monitoring process through Gaussian measurements, the SME preserves the Gaussianity of any initial state. In this case, the vector of average momenta $\bar{\xb}=\langle \hat{\xb} \rangle$ and the Covariance Matrix (CM) $\sigma_{ij}=\langle\{\hat{\xb}_{i},\hat{\xb}_{j}\}\rangle/2-\langle \hat{\xb}_{i}\rangle\langle \hat{\xb}_{j}\rangle$ of the system describe completely the dynamics,through the equations 
\begin{equation}
\label{RiccatiEquation}
\begin{aligned}
\dot{\sigma}&=A \sigma + \sigma A^T + D - \chi(\sigma), \\
\D\bar{\xb}&=(A\bar{\xb}+\mathbf{b})\D t+(\sigma C^T+\Gamma^T)\D\mathbf{w},
\end{aligned}
\end{equation}
where $\D\mathbf{w}$ is a vector of Wiener increments, $A$ $(D)$ is the drift (diffusion) matrix characterizing the unconditional open dynamics of the system, and $\chi(\sigma)$
is a $\sigma$-dependent term that accounts fully for the measurement process. 
Notwithstanding the stochasticity of the overall dynamics, the equation for the covariance matrix is deterministic. 

This is translated in the following expression for the Wigner entropy $S=-\int W\ln{W}\,\D^{2n}\xb$ of the state of the system, which we adopt as our entropic measure~\cite{Adesso2012}
\begin{equation}
S=-\ln({\cal P})+k=
\frac12\ln(\det\sigma)+k,
\end{equation}
with $k$ an inessential constant and ${\cal P}$ the purity of the state of the system, which for a Gaussian state reads ${\cal P}=(\det\sigma)^{-1/2}$. Therefore, $S$ coincides (modulo the constant $k_n$) with the R\'enyi-2 entropy, tends to the von-Neumann entropy in the classical limit~\cite{Santos}, and is a fully deterministic quantity despite the continuous-monitoring process.

As the Wigner entropy only depends on the CM of the system, its evolution is deterministic even for continuously measured system. The same then holds true for the {\it entropy rate} 
\begin{equation}
\label{entropy-rate}
    \frac{\D S}{\D t}=\frac{1}{2}\frac{\D}{\D t}(\rm{Tr}[\log\sigma])=\frac{1}{2}\rm{Tr}[2A+\sigma^{-1}(D-\chi(\sigma))],
\end{equation}
which, from the study of unconditional non-equilibrium thermodynamics~\cite{RMP} of quantum processes, can be split in two contributions as $\dot{S}_{\rm{uc}}=\Phi_{\rm{uc}}+\Pi_{\rm{uc}}$ with $\Phi_{\rm{uc}}$ ($\Pi_{\rm{uc}}$) the unconditional entropy flux (production) rate. In the conditional case, while a similar splitting is indeed possible~\cite{MP} 
both the entropy flux and production rate are inherently stochastic, as they depend on the  first moments of the quadrature operators. 
 Upon taking the average over the outcomes of the measurement, 
the expression for $\dot{S}$ can be recast into the very elegant form
\begin{equation}\label{excess}
    \dot{S}=\dot{S}_{\!\rm uc}+{I},
\end{equation}
where the term $I=\frac{1}{2}\rm{Tr}[\sigma^{-1}\tilde{D}-\sigma_{\rm{uc}}^{-1}D]$ accounts for the excess entropy production resulting from the measurement process, and it is thus information theoretical in nature.
In turn, 
this enables a similar splitting for the entropy production rate as
\begin{equation}
\label{EPRcuc}
    \Pi_c=\Pi_{\rm{uc}}+I.
\end{equation}
This result -- which holds regardless of the Gaussian nature of the system at hand~\cite{MP2} -- connects the entropy production rate of the unmonitored open system to the akin quantity for the monitored one via the informational term $I$. The second law for the un-monitored system, which reads $\Pi_{\rm{uc}}\geq 0$~\cite{RMP}, is now refined as $\Pi\geq I$, which shows the connection between non-equilibrium thermodynamics and information theory. 
The $I$ term depends explicitly on the measurement strategy implemented to acquire information, which in turn affects the dynamics of the measured system, driving it to different final states. We thus have two different yet related mechanisms that we can exploit here: on one hand, we have the dynamics itself, which drives the evolution of a system with the aim, potentially, to optimise thermodynamic performance. On the other hand, we have the well-known possibility to condition the dynamics of a quantum system through the back-action induced by a measurement process. This clearly offers an exploitable mechanism to effectively drive the open dynamics of a system towards a thermodynamics-based criterion for the choice of the specific protocol to be implemented. Indeed, EPR embodies a cost function whose value one aims to control when optimising the conditional dynamics of the system at hand.
\begin{figure}[t]
\centering
\includegraphics[width=0.9\columnwidth]{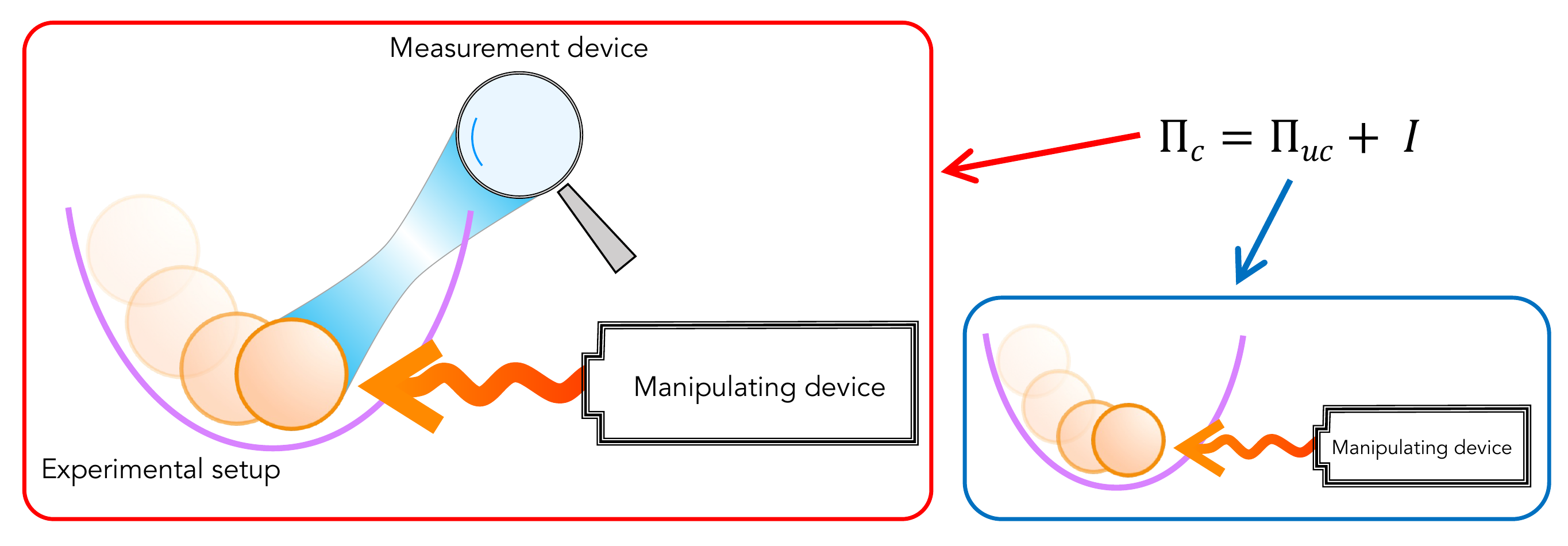}
\caption{Schematic representation describing the difference between the process of unconditional (uc) and conditional (c) dynamics, which, respectively, provide the unconditional $\Pi_{uc}$ and conditional $\Pi_c$ entropy production rate (EPR). Their connection is given by Eq.~\eqref{EPRcuc}. }
\label{fig.EPR}
\end{figure}

\subsection{On Step 4}

We have previously addressed a scheme for feedback cooling of a mechanical system that relied on a closed-loop approach. This is not the only possibility at hand, and one can exploit a different method that, building on the capabilities offered by conditional dynamics ensuing quantum measurements, achieves efficient cooling performances. 

Let us consider a harmonic oscillator of frequency $\omega$, characterized by annihilation and creation operators $a$ and $a^\dag$, respectively, such that $[a,\adag]\!=\!1$. Such bosonic mode or harmonic oscillator comprises a stiffening Duffing-like deformation with strength $\epsilon>0$, such that $\epsilon\ll\omega$, as found in different experimental platforms. In addition, the bosonic mode is coupled to a (spin-like) two-level system via a Jaynes-Cummings interaction.  The Hamiltonian of the system reads 
\begin{align}\label{eq:Hs}
H_{\rm s}=\hbar\frac{\omega_{\rm A}}{2}\sigma_z+\omega\adaga+\lambda(a\sigma^++\adag \sigma^-)
  \end{align}
where $\omega_{\rm A}$ and $\lambda$ denote the Bohr frequency and coupling strength of the two-level system, respectively. We have introduced the spin Pauli matrices, $\sigma_{x,y,z}$ such that $[\sigma_i,\sigma_j]\!=\!2i\delta_{ijk}\sigma_k$ and $\sigma_{z}\!=\!\ket{e}\bra{e}-\ket{g}\bra{g}$ with $\ket{e}$ ($\ket{g}$) the excited (ground) state of the two-level system. Finally, $\sigma^+=(\sigma^-)^\dag=\ket{e}\bra{g}$ is the spin raising operator. 
At the resonant condition $\omega_{\rm A}\!=\!\omega$, the dynamics  governed by the Jaynes-Cummings interaction takes a state $\ket{g,n+1}$ and transforms it into $\ket{e,n}$ in a time $T_n\!=\!\pi/(2\lambda\sqrt{n+1})$ with $n\geq0$.

The goal is to cool an initial thermal state $\rho^{\rm th}_r=\sum_{k=0}p_k\ket{k}\bra{k}$ (with $p_k= n_{\rm th}^k/(1+ n_{\rm th})^{k+1}$ and $n_{\rm th}={\rm Tr}[\adaga \rho_r^{\rm th}]$) of the oscillator down to its ground state by performing measurements on the spin degree of freedom. That is
\begin{equation}
\rho^{\rm th}_r\rightarrow |\psi_{\rm gs} \rangle \langle \psi_{\rm gs}|\approx \ket{0}\bra{0}.
\end{equation}

The model in Eq.~(\ref{eq:Hs}) 
can be realized  in a number of different platforms. Among them, levitated nanoparticles~\cite{Fonseca,Setter}, trapped ions~\cite{Home:11}, circuit quantum electrodynamics~\cite{Ong:11}, and optomechanical systems. 
In the latter context, a spin defect of frequency $\omega_{\rm A}\in [0.5,1.5]\omega$ coupled to a mechanical resonator, $\omega\approx 200$ MHz, and $Q\approx 10^6$, can achieve coupling strengths as large as $\lambda \approx 0.05 \omega$ at negligible spin damping rates~\cite{Tian}. 

The cooling schemes put forward in Refs.~\cite{Li:11, ROM} consider the concatenation of free evolution following Eq.~\eqref{eq:Hs} and projective measurements on the spin degrees of freedom. While Ref.~\cite{Li:11} uses random detection times, the approach of Ref.~\cite{ROM} implies measuring the spin at regular intervals.  
When $N_{\rm rep}$ cycles are implemented, starting from the initial state 
\begin{align}\label{eq:rho0}
\rho_{s}(0)=\ket{g}\bra{g}\otimes \rho_{r}^{\rm th},
  \end{align}
and when the free evolution  in each cycle takes place for a time $T_n\!=\!\pi/(2\lambda\sqrt{n+1})$, we 
remove excitations and entropy, and thus cool down the oscillator. 
After a total time 
\begin{equation}
T_f
=\frac{\pi}{2\lambda} \sum_{n=0}^{N_{\rm rep}-1}\frac{1}{\sqrt{n+1}},
\end{equation}
where a negligible detection time was assumed. The probability of a successful detection of the spin in its ground state $\ket{g}$ upon the evolution $U(T_n)$ is given by $p_{g;n}\!=\!{\rm Tr}[\ket{g}\bra{g}\rho(T_n)]$, which is lower bounded by the probability $p_0=(1+n _{\rm th})^{-1}$ to find the oscillator in its ground state.
 Upon $N_{\rm rep}$ repetitions, success is achieved for  $\Pi_{n=0}^{N_{\rm rep}-1}p_{g;n}\approx p_0$, which turns out to be achievable for $N_{\rm rep}\gg 1$. Hence, one can already notice that this method can be favourable to cool down states of a resonator containing few excitations. In particular, if $ n_{\rm th}\lesssim 10$, we have $p_{0}\gtrsim 1/10$ with $p_k\lesssim 10^{-3}$ for $k\gtrsim 50$, so that $N_{\rm rep}\lesssim 50$ would be sufficient to achieve a significant reduction on the occupation number. However, it is thermodynamically impossible to achieve the ground state of a quantum system in a finite time in light of the third law of thermodynamics and the unattainability principle~\cite{Masanes}, while being still possible to get very close to the actual ground state. 

Remarkably, this approach is effective in cooling also oscillators that showcase a small degree of nonlinearity~\cite{ROM}. For instance, for a quartic potential of the Duffing type $\epsilon(a+\adag)^4$, despite the coupling between states in the Jaynes-Cummings spectrum that do not have the same energy [something impossible to achieve through Eq.~(\ref{eq:Hs}) only], similar cooling performances can be achieved, although oscillators with large values of $\epsilon$ require longer concatenation cycles.

\section{Data intensive tools for quantum computing science}

Quantum computing and artificial intelligence are both transformational technologies in need of each other to achieve significant progress. Although artificial intelligence produces functional applications with classical computers, it is limited by the computational capabilities of the latter. Quantum computing can provide a computation boost to artificial intelligence, enabling it to tackle more complex problems. The visionary goal from the computer science perspective is the achievement of artificial general intelligence, namely the engineering of a system capable of human-level thinking, continuously improving its performances and reaching far beyond current (classical) computational capabilities. For quantum computing, instead, the interplay with artificial intelligence, specifically through machine learning methods, holds the promises to exploit the computational advantages of quantum processors so as to achieve results that are not possible with classical ones. This embodies a new and exciting way to combine a special form of fundamental science, this time deeply rooted into the mathematical and statistics-based efforts aimed at the development of sophisticated algorithmic methods for the processing of large dataset, and quantum technologies for computation and information processing at large~\cite{Carleo}.

\begin{figure}[h!]
\centering
\includegraphics[width=0.9\columnwidth]{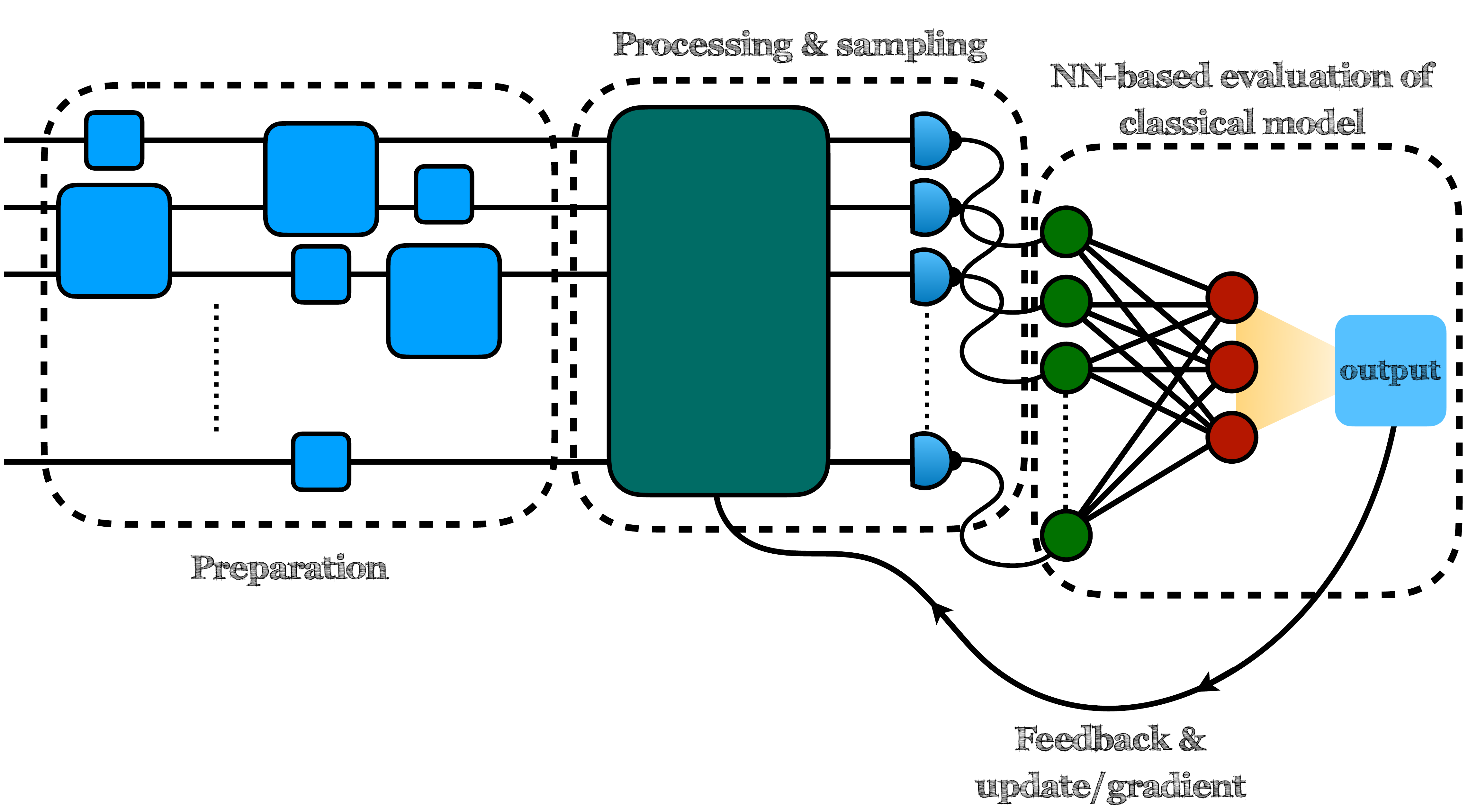}
\caption{General architecture for a machine learning-assisted quantum computational process: the outputs of the quantum processing stage is harvested through a sampling and fed into an NN-based architecture for the evaluation of a classical cost function, whose value informs a feedback loop that changes the parameters characterising the quantum circuit, aiming at  optimisation.}
\label{ML}
\end{figure}

In a typical implementation of a machine learning-assisted quantum computational problem, quantum data obtained from a suitable preparation stage is processed and manipulated by a quantum circuit, implementing the quantum computational stage of the process. Classical information is harvested through a sampling stage where the outputs state of the circuit is probed by an array of quantum measurements. Such classical information is then fed into a classical processing architecture aimed at evaluating a classical model -- for instance through a neural network (NN)-like implementation -- that results in the evaluation of a cost function whose value informs a feedback/updating stage of of the quantum circuit, aimed at optimising the quantum processing steps till convergence is reached [cf. Fig.~\ref{ML}].  The typical complexity of quantum data (stemming from quantum features such as superposition or entanglement) and their fragility to classical and quantum noise make it useful to deploy machine learning-based methods for their interpretation and manipulation.

Crudely, the following embodies a short list of applications of machine learning to quantum information processing problems that are currently being pursued to drive the design of more process-efficient quantum technology~\cite{reviews}
\begin{itemize}
\item    
Development of quantum algorithms for quantum generalization of classical learning models. The scope is the achievement of quantum-induced speed-ups in the deep learning training process, for instance through the fast identification of the optimal values of the weights and links in a NN architecture.
 \item   Quantum algorithms for optimal classical decision problems. The formulation of optimal decision trees is in general complex and cannot be efficiently address by dichotomic approaches leading to classical random walks. Quantum walks, through the coherence-induced fast exploration of configuration space, could allow for the identification of decision  by a number of optimal paths through decision trees faster than any classical walk-based schemes. Such approaches have recently inspired promising techniques for quantum state engineering~\cite{majury}. Quantum efficient searching approaches of this type hold the promise for near-term applications such as efficient data encryption.
\item    Quantum game theory: The systematic extension of classical game theory,  which is widely used in artificial intelligence applications, to the quantum domain will be useful to overcome critical problems in quantum communication and the implementation of a framework for quantum artificial intelligence.
\item Quantum Machine Learning to Solve Linear Algebraic Problems: Many  Data Analysis and machine learning problems are solved by performing matrix operation on vectors in a high dimensional vector space. Quantum computers can solve common linear algebraic problems such as the Fourier Transformation, finding eigenvectors and eigenvalues, and solving linear sets of equations over vector spaces in time that is polynomial in the dimension of the space (and exponentially faster than classical computers due to the quantum speedup). One of the examples is the Harrow, Hassidim, and Lloyd (HHL) algorithm~\cite{HHL}.
\item Quantum Principal Component Analysis: Principal Component Analysis is a dimensionality reduction technique that is used when managing large datasets. Dimensionality reduction comes at the cost of accuracy, as one needs to decide which variables to eliminate without losing important information. 
Classically, dealing with such request efficiently is hard: over large input datasets with many features and attributes, classical methods of principal component analysis will fail because it will be hard for us to visualize the importance of each variable.

Another issue with classical computers is the calculation of eigenvectors and eigenvalues, whose number grows with the dimensionality of the input. Quantum computers can solve this problem very efficiently and at a very high speed by using Quantum Random Access Memory (QRAM)~\cite{QRAM} to choose a data vector at random. It maps that vector into a quantum state. The summarized vector that we get after Quantum Principal Component Analysis has logarithmic size in the number of qubits involved. 
By repeatedly sampling the data and using a trick called density matrix exponentiation, combined with the quantum phase estimation algorithm (which calculates the eigenvectors and eigenvalues of the matrices), we can take the quantum version of any data vector and decompose it into its principal components. Both the computational complexity and time complexity is thus reduced exponentially.

\item Quantum Support Vector Machines: Support Vector Machine is a classical machine learning algorithm used both for classification and regression. For classification tasks it is used to classify linearly separable datasets into their respective classes. Suppose, if the data is not linearly separable, then it’s dimensions are increased till it is linearly separable. 
Quantum computers can perform Support Vector Algorithm at an exponentially faster rate owing to superposition and entanglement.

\item Quantum Optimization:  Optimization is used in a machine learning model to improve the learning process so that it can provide the most adequate and accurate estimations. The main aim of optimization is to minimize a loss function. A more considerable loss function means there will be more unreliable and less accurate outputs, which can be costly and lead to wrong estimations. Most methods in machine learning require iterative optimization of their performance. Quantum optimization algorithms suggest improvement in solving optimization problems in machine learning through the use of multi-party superpositions.

\item Deep Quantum Learning: Quantum computing can be combined with deep learning to reduce the time required to train a neural network. By this method, we can introduce a new framework for deep learning and performing underlying optimization. We can mimic classical deep learning algorithms on an actual, real-world quantum computer. When multi-layer perceptron architectures are implemented, the computational complexity increases as the number of neurons increases. Dedicated GPU clusters can be used to improve the performance, significantly reducing training time. However, even this will increase when compared with quantum computers. Quantum computers  are designed in such a way that the hardware can mimic the neural network instead of the software used in classical computers. Here, a qubit acts as a neuron that constitutes the basic unit of a NN. Thus, a quantum network  can act as an NN and can be used for deep learning applications at a rate that surpasses any classical machine learning algorithm.

\end{itemize}

On the other hand, a number of applications of classical machine learning are driving the development of significantly disruptive methods aimed at demonstrating prototype quantum sensors able to optimally extract information about its environment, achieve fully automatic calibration and operation of multi-qubit circuits and  improve the performance of quantum algorithms for quantum chemistry.

\section{Conclusions}
The translation of research performed in the domain of fundamental science to the development of quantum devices is, likely, the single challenge that the community working in quantum technologies will be called to face as we transition to a phase of consolidation of the field and area. Its successful accomplishment will require a multi-disciplinary approach that tackles aspects that have been, so far, only partially addressed but that are however central to the success of our endeavours in building a quantum technological framework and infrastructure. This contribution has focused on two of such aspects, namely energetics at the fundamental quantum level and the interplay between artificial intelligence and quantum dynamics as two tantalising pathway to explore while pursuing such tantalising ultimate goal. 

\section*{Acknowledgements}
We acknowledge support from the European Union's Horizon 2020 FET-Open project TEQ (766900), the Leverhulme Trust Research Project Grant UltraQuTe (grant RGP-2018-266), the Royal Society Wolfson Fellowship (RSWF/R3/183013), the UK EPSRC (grant EP/T028424/1) and the Department for the Economy Northern Ireland under the US-Ireland R\&D Partnership Programme.

\bibliography{references}

\bibliographystyle{unsrt}

\end{document}